\documentclass[fontsize=10pt, paper=a4, parskip=half]{scrartcl}
\usepackage[T1]{fontenc}

\usepackage[sfdefault,sb]{plex-sans}
\usepackage{plex-mono}

\usepackage{graphicx}
\NewDocumentCommand\emoji{}{\raisebox{-.2\height}{\includegraphics[width=12pt]{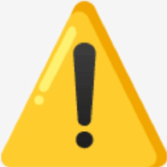}}}

\usepackage{xcolor}

\usepackage[colorlinks=true,allcolors=blue,pdfborder={0 0 0}]{hyperref}
\usepackage{enumitem}

\usepackage[a4paper, left=4cm, right=4cm, top=2.5cm, bottom=2.5cm]{geometry}

\usepackage{fancyhdr}
\usepackage{biblatex}
\setkomafont{title}{\sffamily\Large\bfseries}
\setkomafont{author}{\normalfont\sffamily}
\setkomafont{date}{\normalfont\sffamily}

\usepackage{sectsty}
\sectionfont{\normalfont\sffamily\bfseries\fontsize{14pt}{14pt}\selectfont}
\subsectionfont{\normalfont\sffamily\bfseries\fontsize{10pt}{10pt}\selectfont}

\usepackage{enumitem}
\usepackage{mathabx}
\setlistdepth{8}
\setlist[itemize,1]{label=\textbullet}
\setlist[itemize,2]{label=$\blacktriangleright$}
\setlist[itemize,3]{label=\textopenbullet}
\setlist[itemize,4]{label=$\smalltriangleright$}
\setlist[itemize,5]{label=$\sqbullet$}
\setlist[itemize,6]{label=$\square$=}
\setlist[itemize,7]{label=$\blackdiamond$}
\setlist[itemize,8]{label=$\diamond$}
\renewlist{itemize}{itemize}{8}

\usepackage{tocloft}

\usepackage{setspace}

\usepackage{tcolorbox}
\definecolor{sub}{HTML}{f6f6f6}
\tcbset{
    sharp corners,
    colback = white,
    before skip = 0.2cm,    %
    after skip = 0.5cm      %
}
\newtcolorbox{note}{
    colback = sub,
    colframe = sub,
    boxrule = 0pt,
    left = 33pt,
    right = 8pt,
    top = 8pt,
    bottom = 8pt
}

\usepackage{subcaption}

\title{\texorpdfstring{
    Views on AI aren't binary — they're plural
    \vspace{2mm}
}{
    Views on AI aren't binary — they're plural
}}
\author{{\bfseries Thorin Bristow}\footnote{Independent Researcher (mail@thorinbristow.com)} \and {\bfseries Luke Thorburn}\footnote{King's College London (luke.thorburn@kcl.ac.uk)} \and {\bfseries Diana Acosta-Navas}\footnote{Loyola University Chicago (dacostanavas@luc.edu)}}

\date{\vspace{3mm}September 2024}

\begin{document}

\maketitle
\vspace{-10mm}
\renewcommand\contentsname{}
\tableofcontents
\thispagestyle{empty}
\vspace{8mm}

    \noindent
    In March 2023, the Future of Life Institute published an open letter calling for all AI labs to ``pause for at least 6 months the training of AI systems more powerful than GPT-4'' \cite{fli2023}. The letter generated significant media attention \cite{metz2023,yudkowsky2023,paul2023,loizos2023,bobrow2023} and over 30,000 signatories. Perhaps counter-intuitively, the letter also prompted deeply-felt criticism from another group of researchers who are working to mitigate the societal risks of AI. These researchers argued that the letter placed disproportionate attention on speculative risks, disregarding harms that the AI industry is already causing \cite{luccioni2023,kapoor2023,tante2023,gebru2023,coldewey2023}.

    This exchange reflects an apparent conflict between two groups: those primarily concerned with the immediate harms caused by AI, and those more concerned with the potential future risks of powerful AI systems. These tensions have been present for several years \cite{cave2019}, but became more salient with prominent media attention (e.g., \cite{brauner2023,clarke2023,cowen2023,piper2023,schechner2023}), open letters \cite{cais2023a,faact2023}, and social media posts (e.g., \cite{dobbe2023,lecun2023,whittlestone2023}) throughout the past eighteen months. While the overall debate has evolved and new narratives emerged \cite{welsh2023,zuckerman2024,asparouhova2024}, the perceived conflict between ``AI Ethics'' and ``AI Safety'' continues to influence how debates over AI are framed \cite{bordelon2024,goldman2024,mcbride2024,arnold2024,wang2024} and the extent to which people working on AI development and governance are able to trust each other and collaborate \cite{r7}. While this dynamic points to important differences in approaches to responsible AI, it is all too easy to slip into thinking in binary, us-and-them terms.

    Recent policy debates over climate change and the SARS-CoV-2 pandemic demonstrate how division and politicization over societal challenges can undermine our collective response. The shoehorning of opinions into mutually exclusive groups --- sometimes called \textit{ideological sorting} --- has long been recognized by political scientists and conflict scholars as a risk factor for the escalation of destructive conflict \cite{selway2011,gubler2012,siroky2016}. High levels of sorting make it easier to stereotype and pigeonhole people and reduce representation of nuanced, cross-cutting positions \cite{fiorina2017}. Conversely, low levels of sorting increase the number of ``surprising validators'' who help make groups legible to one another \cite{glaeser2014}, and increase the likelihood that any majority will include representatives of any minority, reducing the risk of majoritarian tyranny and the formation of epistemically flawed echo chambers. Allowing sorted, us-vs-them conflict --- \textit{high conflict} \cite{ripley2022} --- to emerge in the broad community of people working on responsible AI would hinder our ability to ensure that the impacts of emerging AI are inclusively beneficial \cite{caviola2024}.

    With this in mind, our aim for this piece is to support those navigating the societal and political issues related to AI by providing context and language with which to understand these two perspectives, the apparent conflict between them, and the broader intellectual environment in which they emerge. Specifically, we (i) describe the emergence of this false binary, (ii) explain why the seemingly clean distinctions drawn between these two groups don’t hold up under scrutiny, and (iii) provide concrete suggestions for how individuals can help prevent polarization and politicization of AI discourse.

    Our interest in this paper is to address a \textit{meta-level} concern regarding plurality in the field of responsible AI. As such, while we provide references for further consultation and a \hyperref[sec:glossary]{glossary} of key terms, we refrain from substantive discussion of the ethical issues that are the ordinary focus of this field.

\section{The false binary}

    \begin{note}\raggedright
        \hspace{-25pt}\emoji\hspace{13pt}\textbf{Note to readers}\\[2pt]
        If you are short on time, please skip this section (where we describe a stereotype) and go straight to the \hyperref[sec:complex-reality]{next section} (where we debunk it). The debunking is more important!
    \end{note}

\subsection{The caricature}

    \begin{itemize}
        \item The caricatured view of this conflict is one of binary camps, one most concerned with ``near term'' social harms that are already occurring, and the other concerned with ``long term'' harms from ``misaligned'' AI, possibly including the risk of human extinction.
        \item We emphasize that we describe these stereotypes only because they are operative in public debates over AI, contribute to the tensions which are the focus of this paper, and so merit examination. We don't endorse their use.
    \end{itemize}
    
\subsection{A note on language}

    \begin{itemize}
        \item In fact, we would like to begin by acknowledging that the very language used in these discussions is fraught.
        \item There are failures of accuracy, and unacknowledged complex moral implications, in some of the common terms used to frame this debate:
        \begin{itemize}
            \item ``short term'' / ``near term''
            \begin{itemize}
                \item is inaccurate because most of the issues to which it refers (bias, inequality, discrimination, etc.) will be around for the foreseeable future \cite{critch2023}, and
                \item may be used in a belittling way, insofar as it could suggest that those who regard these issues as high-priority are, for that reason, myopic.
            \end{itemize}
            \item ``long term''
            \begin{itemize}
                \item is likely inaccurate because concrete manifestations of the control problem (the basis of many ``long term'' concerns, see the \hyperref[sec:glossary]{glossary}) may cause significant harm within the next decade, if not already \cite{critch2023}, and
                \item can be construed as self-congratulatory by implying farsightedness on the part of those who identify with it.
            \end{itemize}
            \item ``ethics''
            \begin{itemize}
                \item is misleading, because the ideas of ``longtermism'' and ``effective altruism'' are themselves views in moral philosophy (e.g., the protection of human well-being and interests in the distant future is itself an ethical question), and
                \item is often interpreted shallowly as referring to a mere list of issues in AI models (bias, fairness, etc.) that doesn't reflect the breadth and nuance of thinking in the field of moral philosophy and its applications to technology.
            \end{itemize}
            \item ``safety''
            \begin{itemize}
                \item is misleading, because using it exclusively to refer to risks of human extinction implies that AI-facilitated biases or discrimination --- for example, in law enforcement settings --- are not matters of safety for those affected \cite{dobbe2023,wang2024}.
            \end{itemize}
            \clearpage
            \item ``alignment'' / ``misalignment''
            \begin{itemize}
                \item can be used as a euphemism to describe trade-offs and harms that would otherwise be spoken about in more confronting terms (e.g., it seems euphemistic to refer to biased automated decision-making that effects someone's hiring prospects, credit score or criminal record as a mere ``misalignment''), and
                \item implies a simplistic intuition of two well-defined things being made equal or flush, which is not an accurate model for many of the problems it is used to describe \cite{grietzer2024}. For example, it glosses over
                \begin{itemize}
                    \item ambiguity in whether AI should align with human goals, interests, expressed or revealed preferences, values, capabilities, or some other notion of preference \cite{gabriel2020}, when these alternative notions can differ \cite{barocas2014,zhixuan2024},
                    \item the decision of to what extent AI should align to individual or collective preferences \cite{kirk2024},
                    \item the many mathematical challenges of aggregating individual preferences into coherent notion of collective interest \cite{maskin2014},
                    \item the fact that disadvantaged individuals can learn to lower their expectations even if it is not in their best interests to do so, a phenomenon known as ``preference adaptation'' \cite{khader2011,sen1999},
                    \item the many cases in which there are ``hard problems'' \cite{chang2017} -- difficult and inevitable trade-offs among competing preferences \cite{bondi2021}, and
                    \item the fact that an AI system which is ``aligned'' in one context may not be aligned in another.
                \end{itemize}                
            \end{itemize}
        \end{itemize}
        \item Despite these terminological imperfections, it is important to discuss these stereotypes as they operate in public debate. For this reason, we use capitalized Ethics and Alignment as shortcuts to refer to them.
        \item We provide a \hyperref[sec:glossary]{glossary} of key terms in the appendix.
    \end{itemize}

\subsection{Ethics {\normalfont(the stereotype)}}

    \begin{itemize}
        \item Sometimes referred to using the terms ``FAccT'' (Fairness, Accountability \& Transparency), ``FATE'' (Fairness, Accountability, Transparency \& Ethics), ``AI ethics,'' ``near term,'' or ``short term.''
        \item According to the stereotype, Ethics
        \begin{itemize}
            \item cares mainly about social harms, which are often the result of systemic and structural factors and are not captured by existing legal frameworks, such as those related to inequality, bias, fairness, power imbalances, mistreatment of workers, and environmental sustainability \cite{pemberton2016},
            \item views the risk of social harms stemming from AI as symptomatic of pervasive human prejudices and structural inequalities that are reflected and amplified by technology,
            \item regards the root of the problems as social in nature,
            \item rejects recent AI hype as unjustified,
            \item holds that those involved in the development of AI have a high degree of agency over its impact and should bear responsibility and accountability for the technology they build, and
            \item rejects the philosophical ideas of Effective Altruism and longtermism that inform Alignment.
        \end{itemize}
    \end{itemize}

\subsection{Alignment {\normalfont(the stereotype)}}

    \begin{itemize}
        \item Sometimes referred to using the terms ``AI safety'' or ``long term.''
        \item According to the stereotype, Alignment
        \begin{itemize}
            \item cares mainly about the risk of human extinction brought about by a powerful AI whose goals are not attuned with human interests  (``existential risk'' or ``x risk''),
            \item views this risk mainly through the lens of the ``alignment problem'' or ``control problem'' (the difficulty of perfectly specifying nuanced and complex human objectives in code),
            \item regards the problems as technical in nature,
            \item holds that recent AI hype is justified,
            \item defaults to the idea that, while developers have some agency over the impact of AI, the future is partly determined by geopolitical factors, economic and game theoretic dynamics which are out of individuals' control, and
            \item is motivated by Effective Altruism and longtermism \cite{macaskill2015,macaskill2022,singer2015}.
        \end{itemize}
    \end{itemize}     

\subsection{Ethics's discontents with Alignment}

    \begin{itemize}

        \item \textit{Flawed empirical bases}
        \begin{itemize}
            \item Ethics is skeptical of the plausibility and immediacy of the risks described by Alignment. They reject Alignment's assumptions about the capabilities of the technology and regard it as ``hype'' \cite{westerstrand2024,narayanan2024a}. They also criticize Alignment for not describing \textit{how} AI poses an existential risk \cite{cristianini2023,lazar2023a,thorstad2024}, for assigning probabilities to catastrophic scenarios without evidence \cite{narayanan2024}, for drawing faulty analogies between AI and nuclear weapons \cite{kaushik2023}, for fabricating a story about an AI seeking to eliminate its operator \cite{kaushik2023}, and for making absolute claims, without acknowledging uncertainty \cite{richards2023}.
        \end{itemize}

        \clearpage
        \item \textit{Homogeneity and blind spots}
        \begin{itemize}
            \item Ethics criticizes Alignment as being a cloistered group without meaningful sociological diversity. This raises a concern that researchers in Alignment predominantly hail from a limited range of educational backgrounds, elite educational institutions, and from specific demographic profiles. This homogeneity renders them ill-equipped to represent the perspectives of broader society. While Alignment frames its priorities as maximally impartial and focused on the general interest of humanity, Ethics argues this view is inevitably shaped by the group's homogeneity, predominantly consisting of highly-educated white men from relatively affluent neo-atheist/ secular backgrounds. This insularity raises suspicion from Ethics about whose values and experiences underpin the problems/solutions proposed by Alignment, specifically their view that key sociological considerations may be overlooked or dismissed as irrelevant. The uniformity of the group, hold Ethics advocates, renders them unable to appreciate the importance and urgency of such considerations \cite{lazar2023a,ahmed2023}.
        \end{itemize}
        
        \item \textit{Failures in integrity}
        \begin{itemize}
            \item Ethics criticizes Alignment for being elitist. They hold that Alignment's focus on existential risk can be traced back to a lack of care for algorithmic harms that will likely be concentrated in minority groups \cite{hine2023}, as opposed to affecting the majority of humanity. The focus on existential risk among elites is thought to be explained by the fact that elites are ``unlikely to experience harm from `AI' unless everyone else on the planet does first'' \cite{bender2023}.
            \item Researchers in Ethics have raised concerns related to the existence of a flawed ideology at the core of Alignment's research agenda. Some have claimed all Alignment concerns are fundamentally rooted in eugenics \cite{bender2023}, and coined the acronym ``TESCREAL'' (Transhumanism, Extropianism, Singularitanism, Rationalism, Cosmism, EA, Longtermism) which is used to jointly denote a set of distinct but overlapping communities from whose ideas the Alignment perspective emerged, specifically ideas about the significance of pursuing ``artificial general intelligence'' (AGI) and the role they see for AGI in human history \cite{gebru2024,gebru2023a,linton2023,torres2023,torres2023a,torres2024}. Some in Ethics argue that the influence of this set of ideologies undermines the moral and intellectual legitimacy of the Alignment agenda.
            \item Alignment has also been criticized for failing to appropriately deal with significant conflicts of interest, including among political advisors and their attempt to shape policy and legislation to serve the interests of a narrow group \cite{bordelon2023}.
        \end{itemize}
                
        \item \textit{Resource imbalance}
        \begin{itemize}
            \item Ethics is concerned that allocation of resources, both financial and intellectual, towards existential risks diverts attention and funding away from addressing pressing sociological harms. Alignment is viewed to have been more organized, homogeneous and coordinated \cite{bordelon2023}, and more effective at recruiting college students \cite{tiku2023}, getting sympathetic people into influential positions \cite{bordelon2023}, and framing prominent government responses through the lens of Alignment \cite{clarke2023,bordelon2024}. Ethics feels that the practical outcome of the Alignment narrative is to ``hijack'' the conversation \cite{tiku2023}, distracting regulators, academics, and funders from addressing serious current harms, and helping recruit for corporations who actively perpetrate those harms \cite{richards2023,gebru2022,lindsay2023}.
            \item The existence of this resource imbalance is itself sometimes disputed by Alignment \cite{byrnes2023}.
        \end{itemize}

 \item \textit{Maltreatment}
        \begin{itemize}
            \item Ethics has encountered resistance from Alignment, which downplays the significance of the risks identified by Ethics and disregards their expertise in the relevant areas \cite{r6, yudkowsky2024}. For example, their concerns have been dismissed as merely arguments about status \cite{byrnes2023}, or as politically one-sided ``pet issues'' \cite{r7}, less existentially serious than the concerns of Alignment \cite{oneil2023}.
            \item Many who have devoted their careers to raising awareness of algorithmic harms have had their concerns fall on deaf ears, and struggled to attract attention and funding \cite{piper2023}. Some have been professionally penalized for raising these issues \cite{oneil2023}, often by companies that now invest heavily in (or at least pay lip service to) the concerns of Alignment. In the meantime, members of the Alignment community have been observed gloating about the media attention their ideas have received \cite{hwang2023,shavit2023}.
        \end{itemize}
        
    \end{itemize}

\subsection{Alignment's discontents with Ethics}

    \begin{itemize}

        \item \textit{Inadequate consideration of catastrophic scenarios}
        \begin{itemize}
            \item Alignment argues that Ethics dismisses the risks of future uncontrollable AI without having sound arguments for doing so \cite{alexander2021,bengio2024}. Consequently, they worry that solely addressing the algorithmic harms that Ethics focuses on might be analogous to rearranging deck chairs on the Titanic.
            \item Alignment advocates worry that dismissals of catastrophic risks fail to acknowledge (1) that AI capabilities are currently growing at an unpredictable pace; and (2) the current nonexistence of any methods to ensure that the goals of AI are compatible with human collective goals. Absent such methods, they worry that powerful actors may use the power of AI to the detriment of other human beings \cite{bengio2024}.
            \item Alignment claims that Ethics often resorts to pejoratives rather than fully-fleshed out arguments. For example, Ethics has dismissed Alignment's concerns as fearmongering and alarmism \cite{nature2023,bordelon2023,kaushik2023}, as a fetish or fixation \cite{kaushik2023,bordelon2023}, as ``criti-hype'' \cite{tante2023}, as pseudo-religious \cite{richards2023,torres2021,torres2021a}, as being untethered from reality \cite{bordelon2023,bender2023,clarke2023}, as containing nothing of value \cite{bender2023}, and as ``wishful-worries'' --- ``problems that it would be nice to have, in contrast to the actual agonies of the present'' \cite{brock2019}.
            \item Alignment objects that Ethics dismisses recent AI advances (for example, dismissing large language models as ``spicy autocorrect'' \cite{oneil2023}) without seriously reckoning with their potential to form the basis of autonomous artificial agents.
            \item Alignment researchers oppose what they perceive as social and professional punishment for good faith efforts at raising consequential concerns \cite{schechner2023}.
        \end{itemize}
            
        \item \textit{Attacks on integrity}
        \begin{itemize}
            \item Researchers in Alignment have objected against the accusation of being inspired by a corrupt ideology (TESCREALism), describing this criticism as unfair and unjustified \cite{scharmen2024}. Some researchers, both inside and outside Alignment, have argued that such criticism conflates ideas that are actually distinct. They hold, moreover, that the careful consideration and prevention of AI-driven catastrophic scenarios is legitimate and worthwhile, independent of the merits or demerits of other views held by people involved in the development of these ideas \cite{sennesh2023}.
        \end{itemize}
        
    \end{itemize}

\section{The complex reality}
\label{sec:complex-reality}

\subsection{Where Ethics and Alignment (actually) differ}

    \begin{itemize}
        
        \item \textit{Likelihood of rogue agents}
        \begin{itemize}
            \item Arguably one of the root differences between Ethics and Alignment is the credence they assign to the risk of out-of-control AI. Generally, the Alignment community think that if technology continues to develop, eventually it will be possible to build autonomous agents that we will be wholly unable to control. In contrast, Ethics think this outcome is of negligible probability, if not impossible, in the foreseeable future.
            \item This underlying disagreement fuels many downstream debates, including those about
            \begin{itemize}
                \item stakes and urgency,
                \item how resources should be allocated,
                \item the degree to which resources are zero sum \cite{byrnes2023,alexander2021a},
                \item what kinds of (regulatory) interventions are proportionate, and
                \item the degree to which people expressing concern about the risk of rogue agents can be assumed to be acting in good faith.
            \end{itemize}
        \end{itemize}
        
        \item \textit{Values and priorities}
        \begin{itemize}
            \item Independently of the probability they assign to catastrophic outcomes caused by powerful AI, researchers who identify more closely with one or the other stereotype differ in their values and priorities.
            \begin{itemize}
                \item The research agenda of Alignment expresses mainly utilitarian values, seeking to protect the interests of the greatest number of people, including future generations. Their work is oriented towards finding technical solutions to abstractly formulated theoretical problems, as opposed to existing social realities \cite{weinstein2021}.
                \item Ethics's agenda reveals prioritarian or egalitarian values, which direct us to attend the needs and interests of the people who bear the bulk of burdens in the distribution of social goods. Their work is oriented toward mitigating serious ongoing harms to individuals living under conditions of oppression and disadvantage. 
            \end{itemize}
            \item These differences may, in fact, be driven to some extent by differences in backgrounds, and relative privilege of  thought leaders and individual members.
            \item These differences may also account for downstream debates regarding the importance of allocating resources to address one set of concerns as opposed to another, the stakes, and the urgency of doing so. 
        \end{itemize}
        
        \item \textit{Culture and norms}
        \begin{itemize}
            \item Ethics and Alignment, to the extent that they describe coherent communities, come from different academic disciplines and cultures.
            \item Ethics has grown out of work on algorithmic bias and fairness, machine ethics, critical theory, and \textit{Science and Technology Studies} (STS). While the Alignment community has roots in problems of control theory that were articulated in the mid-twentieth century, it is in its current form populated to a large extent by members of a particular epistemic community, who are motivated to mitigate existential risk \cite{ahmed2023}. Each of these communities have distinct cultures and norms \cite{marantz2024}, and in many cases have not substantively interacted, which can lead them to talk past one another and increase the risk of animosity \cite{byrnes2023a}.
        \end{itemize}

        \item \textit{Degrees of agency}
        \begin{itemize}
            \item Sociological problems are much harder to tackle than technical ones, yet Ethics believes in developers’ ability to do something about them. 
            \item Technical problems are easier to tackle and definitely within the scope of developers’ choices, yet Alignment is skeptical about developers' ability to mitigate them. 
            \item This signals an underlying difference in conceptions of the moral agency and responsibility of developers and other actors in the AI ecosystem.
        \end{itemize}
        
    \end{itemize}

\subsection{Where Ethics and Alignment (actually) are similar}

    \begin{itemize}

        \item There is considerable common ground \cite{arnold2024,schuett2023}. Both Ethics and Alignment
        \begin{itemize}
            \item aspire to ethical and responsible development of AI,
            \item are concerned about and interested in addressing the root causes of algorithmic harms, including competitive pressures, limited accountability, and the control problem \cite{brauner2023},
            \item believe that technology companies usually do not take AI ethics and responsibility seriously \cite{piper2023}, 
            \item acknowledge the need for robust regulatory and governance mechanisms, supporting many of the same policy proposals \cite{cais2023,hausenloy2023}, 
            \item recognize that addressing AI risk requires input from many disciplines, stakeholders, and perspectives,
            \item share a concern for potential unintended consequences, emphasizing need for impact assessments, and anticipating and mitigating potential risks and harms, and
            \item appear to share an underlying assumption about the risk of path dependence (that is, a concern that design flaws that are coded into the technology today will be more difficult to reverse in the future), thus raising the probability of harm.    
        \end{itemize}

    \end{itemize}

\subsection{Complication: There are more than two camps}

\begin{figure}
        \centerline{\includegraphics[width=1.2 \textwidth]{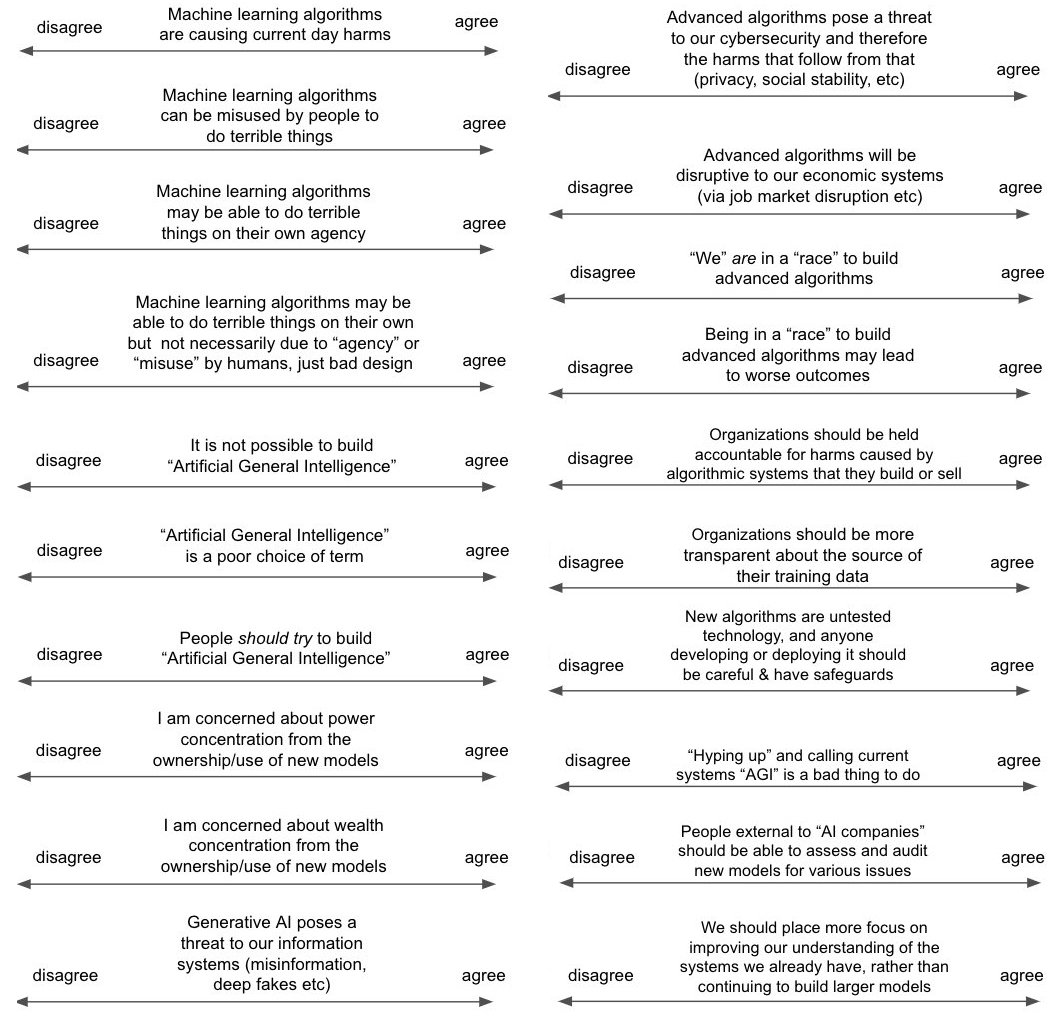}}
        \caption{Examples of the many axes along which views on AI vary. Adapted from \textcite{bluemke2023}.}
        \label{fig:axes}
    \end{figure}

    \begin{itemize}
        
        \item The space of perspectives on AI risk is much more complex than can be accurately described using only two ``camps''. The stereotypes of Ethics and Alignment are not a simple partition of the space: they overlap, disagree internally, and fail to represent many perspectives that agree or disagree with elements of both \cite{richards2023,schechner2023}.
        
        \item There are many fundamental issues about which people disagree \cite{cotra2023,stray2023,nielsen2023}, including
        \begin{itemize}
            \item whether AI systems will soon be able to complete the majority of economically useful tasks that humans currently do,
            \item whether ``intelligence'' can be quantified,
            \item whether recent developments in generative AI represent a significant step change in the capabilities of AI,
            \item whether the concentration of power in AI companies is antidemocratic,
            \item the degree to which we have agency over the trajectory of technologies, or conversely, are victims of outcomes predetermined by game theory \cite{westerstrand2024,nielsen2024},
            \item the existence (and implications) of a global race to build the most advanced AI systems,
            \item whether corporate AI companies benefit from directing attention towards concerns about existential risk,
            \item how much individual control over AI systems is desirable \cite{ge2024}, and
            \item whether open source foundation models will pose a significant risk going forward \cite{bommasani2023,kapoor2024}. 
        \end{itemize}
        
        More axes of disagreement are listed in Figure \ref{fig:axes}. Positions on each of these issues may correlate with the Ethics/Alignment divide, but there are many whose combination of positions (e.g., \cite{cowen2023,andreessen2023}) on these issues means they cannot be accurately described by either stereotype. 

    \end{itemize}

\begin{figure}
        
        \begin{subfigure}{1\textwidth}
            \centerline{\includegraphics[width=1.3\textwidth]{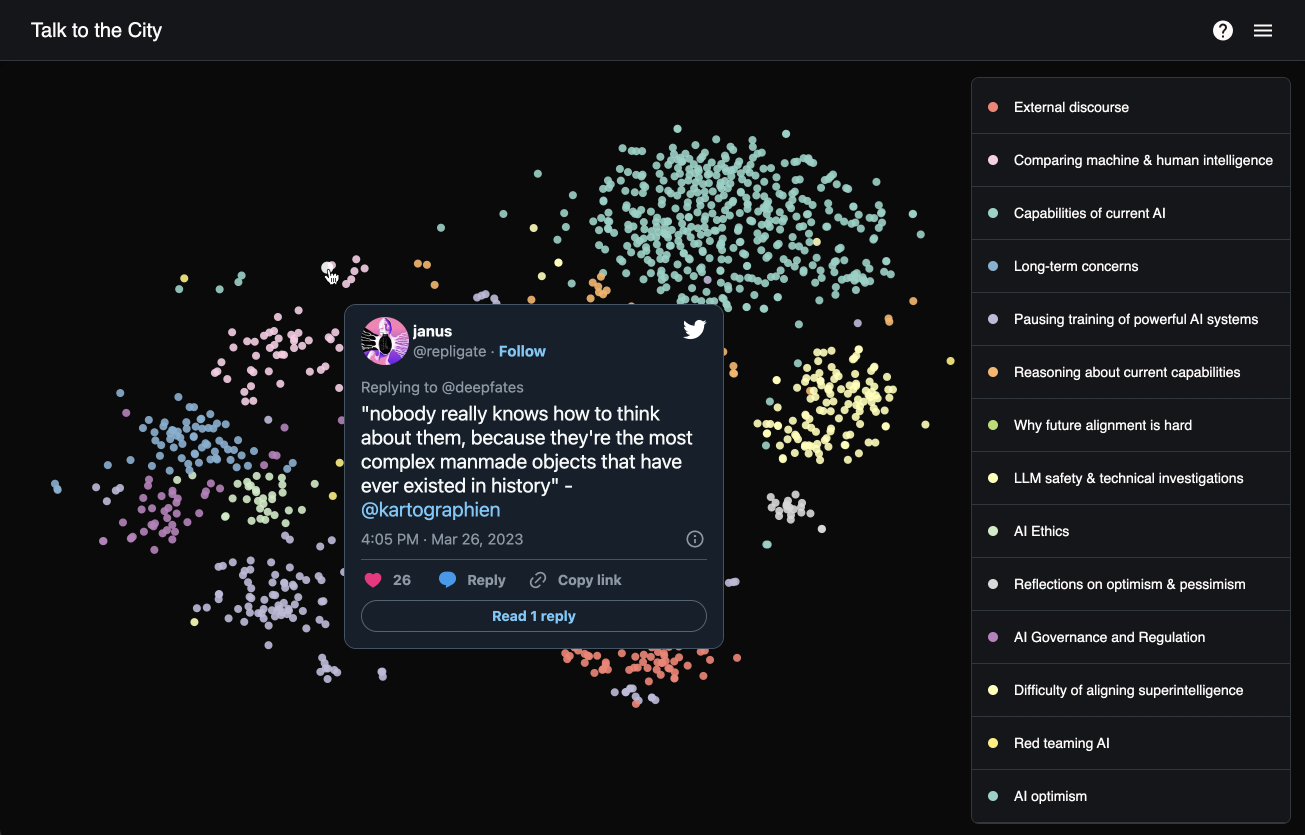}}
            \caption{From \textcite{turan2023}.}
        \end{subfigure}\\ 

        \centerline{
        \begin{subfigure}{0.9\textwidth}
            \includegraphics[width=\textwidth]{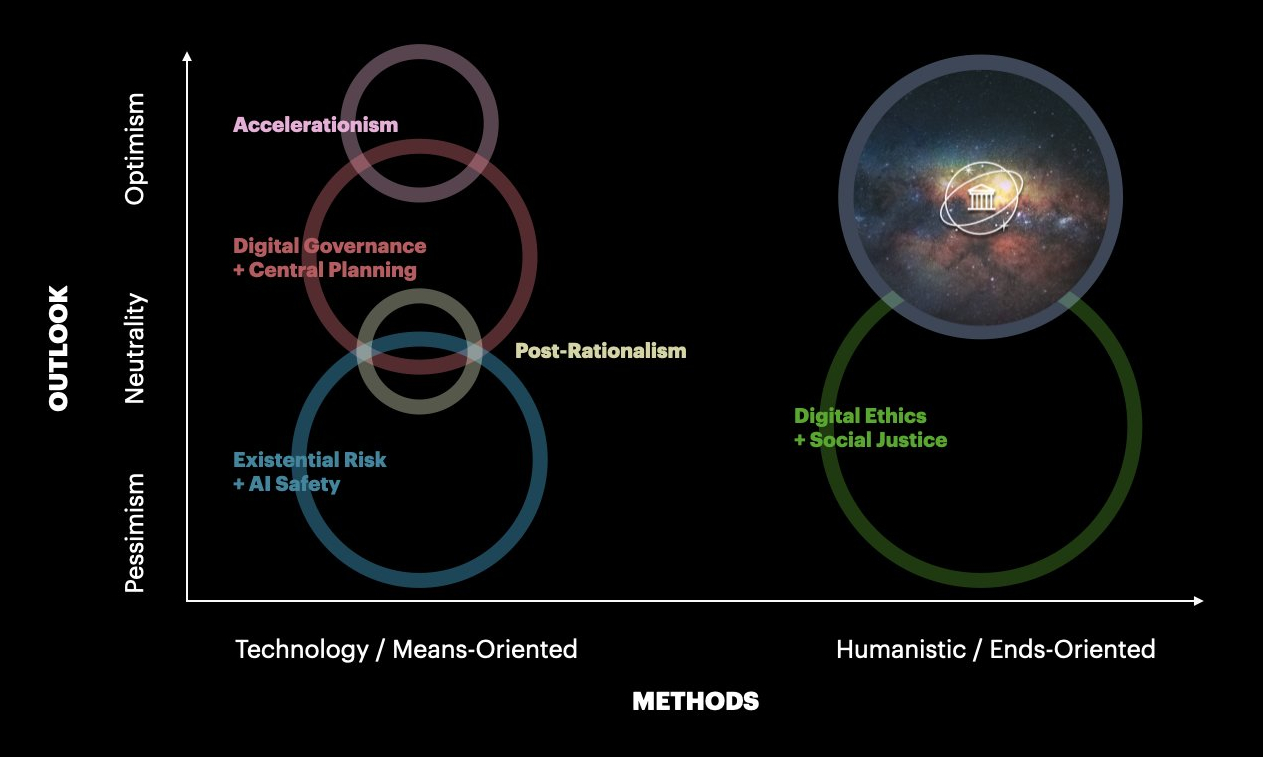}
            \caption{From \textcite{mccord2024}.}
        \end{subfigure}
        \hspace{.03\textwidth}
        \begin{subfigure}{0.475\textwidth}
            \includegraphics[width=\textwidth]{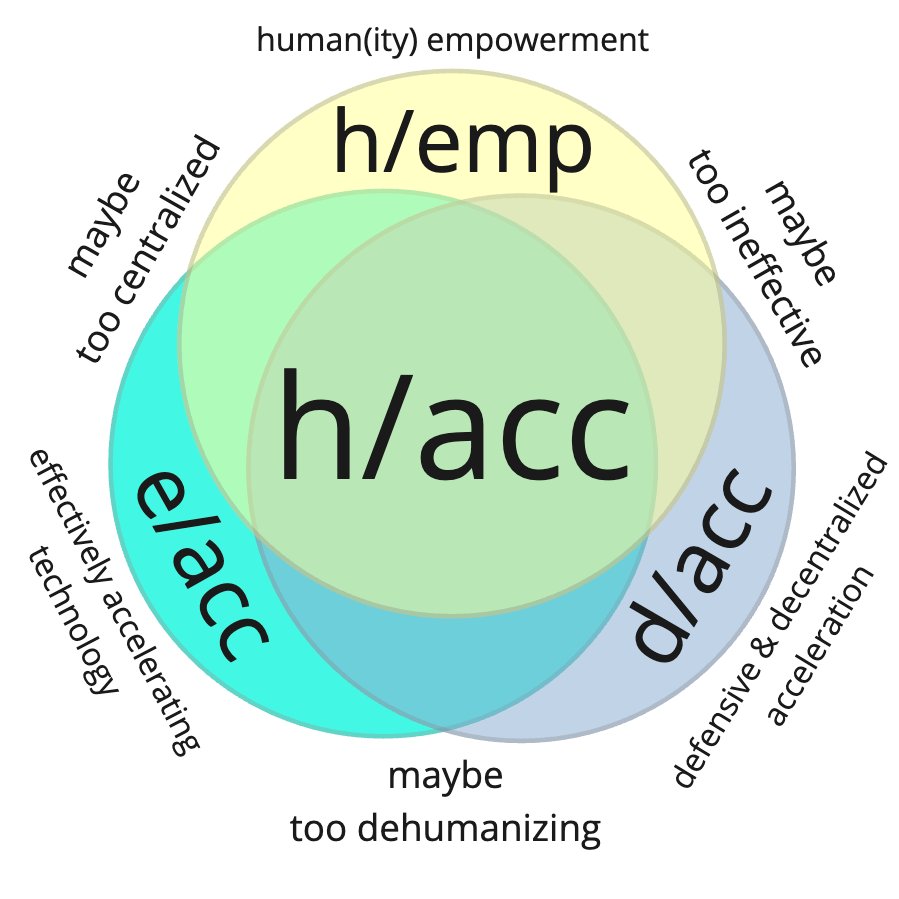}
            \vspace{1pt}
            \caption{From \textcite{critch2024}.}
        \end{subfigure}
        }
            
        \caption{Attempts to map the space of perspectives on AI.}
        \label{fig:models-1}
        
    \end{figure}

\begin{figure}[t]
        \centerline{\includegraphics[width=1.2\textwidth]{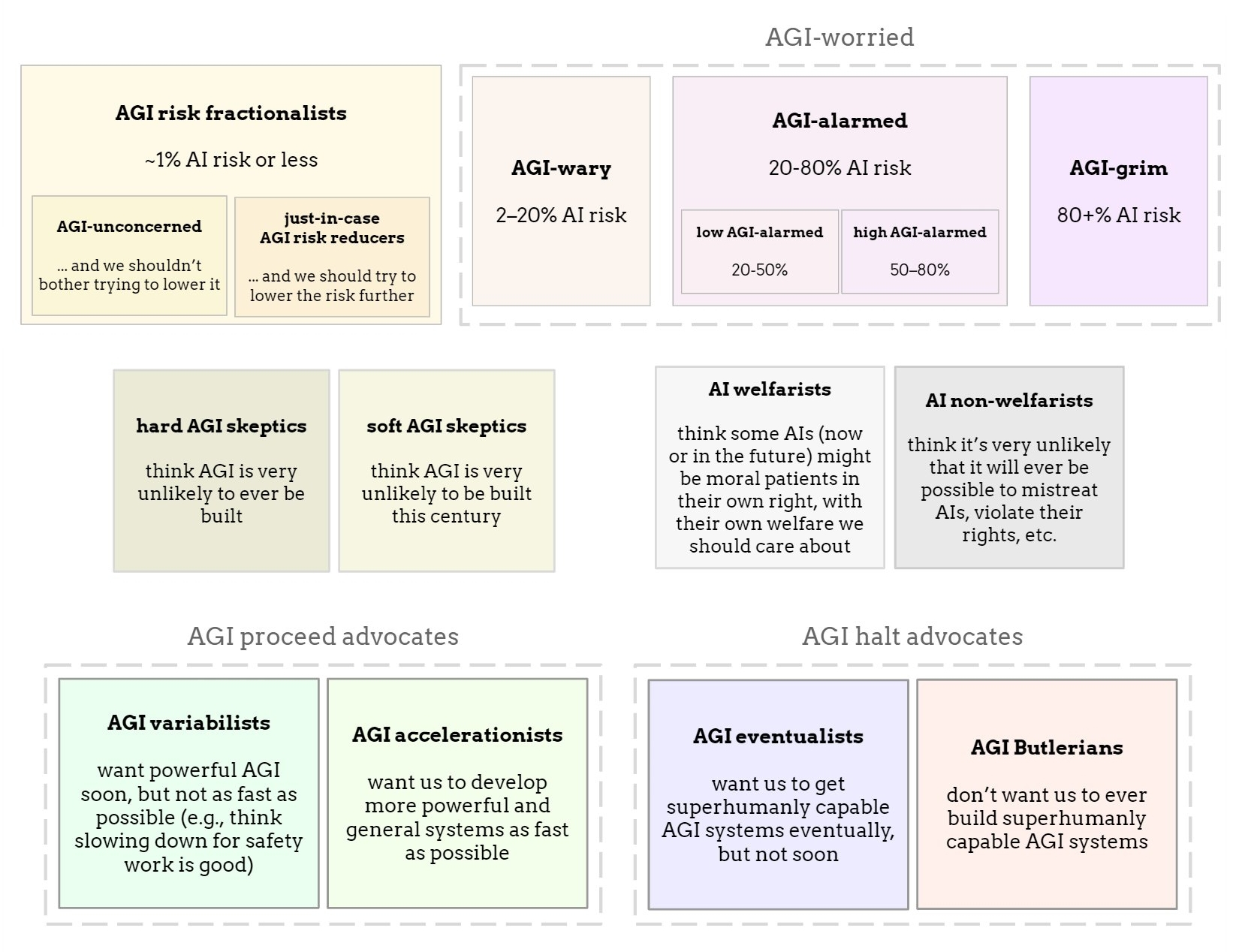}}
        \caption{Another map, from \textcite{bensinger2024}.}
        \label{fig:models-2}
    \end{figure}

    \begin{itemize}

        \item There are many more than two groups. Other prominent identities include ``e/acc'' \cite{jezos2022,welsh2023,wilhelm2023,torres2023b,asparouhova2024}, ``d/acc'' \cite{buterin2023}, AI security professionals \cite{gabe2024}, the majority world \cite{cave2023}, regulators, the VC/startup community \cite{svangel2024}, the open source community \cite{lambert2024}, Big Tech, libertarians, the Web3 community, and many others. We are not aware of any attempts to comprehensively map the space, but Figures \ref{fig:models-1} and \ref{fig:models-2} contain some models that go beyond the false binary.

    \end{itemize}
    
    \begin{itemize}

        \item Even to the limited extent that the Ethics and Alignment identities are coherent, they are cross-cut by political and demographic identities. For example, the risk of uncontrollable AI is taken seriously (and not taken seriously) by some on the political left and right \cite{lovely2022,mein2023,byrnes2023}, and by members of many demographic groups \cite{elsey2023}. Cultural differences also shape expectations around AI systems in ways that are not aligned with the binary division \cite{ge2024}. The existence of such cross-cutting identities is another reason why it is inaccurate to treat Ethics and Alignment as monolithic groups, and signals that --- at present --- this divide hasn't been entirely co-opted by mainstream partisan politics.
    
    \end{itemize}

\subsection{Complication: The existential risk narrative has corporate value}

    \begin{itemize}

        \item Talking about existential risk benefits technology companies. Specifically, it
        \begin{itemize}
            \item helps with PR, by suggesting their technology is powerful and sophisticated \cite{schechner2023,r3},
            \item shapes policy and regulation in their favor, such as by
            \begin{itemize}
                \item garnering influence by allowing ``a homogeneous group of company executives and technologists to dominate the conversation about AI risks and regulation'' \cite{nature2023}, sending a message to regulators along the lines of: ``you should be scared out of your mind, and only we can help you'' \cite{bordelon2023,clark2023},
                \item encouraging burdensome red tape that stymies competition and locks in the advantages of large corporations \cite{bordelon2023,davidson2023,lecun2023,bennett2024},
                \item enabling companies to look like they are calling for regulation, while avoiding regulation of ``lesser” technologies that may not amount to existential risk \cite{bryson2023,r3,wong2023}. ``We’re having conversations about some unknown machine entity destroying the the world ... we’re not talking about humans being underpaid. We’re not talking about data being stolen. We’re not talking about privacy. We’re not talking about safety'' \cite{gebru2023b}, and
                \item drawing attention toward long-term existential risks, thus allowing technology companies to avoid pressing legal and ethical questions, including issues related to copyright and surveillance.
            \end{itemize}
            \item helps with recruitment and talent acquisition, by sounding important and prestigious, and
            \item encourages investment, by supporting the view that AI will be extremely powerful and thus fueling global races for the most powerful AI systems \cite{nature2023}.
        \end{itemize}

        \item Talking about existential risk benefits media companies. Specifically, it
        \begin{itemize}
            \item increases engagement and revenue, by sounding sensational and dramatic, and
            \item increases credibility, because of a widespread bias that views pessimists as intelligent and optimists as naive \cite{ridley2010}.
        \end{itemize}
        
        \item Because of this, well-meaning members of academia and civil society can be unintentionally promoting corporate interests --- and accelerating harms those corporations cause --- when they talk about existential risk from AI \cite{r2}.

    \end{itemize}

\subsection{Complication: Good and bad ideas can be knotted together}
\label{sec:complication-3}

    \begin{itemize}

        \item Speaking generally, valuable and objectionable ideas can be knotted together. For example, a given person can espouse both valuable and objectionable views, and research areas or communities might have objectionable origins but evolve in valuable directions (or vice versa).

        \item The intellectual work of (1) identifying and critiquing flawed elements of particular bodies of thought, and (2) engaging with and building on what is valuable, is critical for helping conversations mature, surfacing good ideas, and moving forward constructively.
        
        \item Wading through the extent and depth of the influence that corrupt ideologies have over the Alignment movement is outside the scope of this piece. However, it should be noted that the very critics of the TESCREAL bundle of ideologies recognize that the degree to which they are endorsed by AI developers and researchers is uncertain \cite{gebru2023a}. It is therefore important to identify which specific aspects of the agenda are subject to this influence and which are not. This is a subject for important future research, as well as something to be mindful of for those entering the field. 
        
        \item A relevant philosophical distinction that may help clear the terrain is the Rawlsian distinction between ``concept'' and ``conception.'' In Rawls's \textit{A Theory of Justice}, a 
        \begin{itemize}
        \item ``concept''  
            \begin{itemize} 
                \item refers to a general, overarching idea or principle that can be widely agreed upon but is relatively abstract and vague,
                \item for example, the concept of justice is universally recognized as important, but its exact definition might vary. 
            \end{itemize}
        \item ``conception'' 
            \begin{itemize}
                \item refers to a specific, detailed interpretation or application of that general concept,
                \item for example, different social groups hold different conceptions of justice, based on how they interpret other concepts like fairness, equality, moral desert, etc \cite{rawls1971}.
            \end{itemize}
        \end{itemize}
        
        \item The ideas that are criticized for falling under the TESCREAL umbrella are part of a specific conception of AI safety, which takes a particular stance on the  place of AGI in human history and its desirability \cite{gebru2023a}. Teasing apart the concept of AI safety (or responsible AI more broadly) from specific, objectionable conceptions of it may help us develop new conceptions that are more well-founded.
        
        \item This kind of work --- deliberating over the relative merit of different conceptions of AI safety, and articulating alternatives and refinements thereof --- calls for a different kind of moral response than questions about the moral impurity of the entire agenda. Among other things, it places greater emphasis on acknowledging good faith where it exists, and offering viable paths to redemption for those whose work is being criticized.

    \end{itemize}

\section{Overcoming the dichotomy}
    
\subsection{Why should we?}

    \begin{itemize}
        \item Many have raised concern about the emergence of the false binary and called for more nuance and efforts to find common ground \cite{heigeartaigh2023,saetra2023,whittlestone2023,richards2023,schechner2023,giansiracusa2023,stix2021,cave2019,mccord2024,johnson2024,arnold2024}. The extent of the conflict is overstated by both traditional and social media, and it is important that people do not let themselves be drawn in to the binary stereotypes and allow them to become a self-fulfilling prophecy \cite{mitchell2023}.

        \item We acknowledge that some have criticized the idea of bridging this divide, or the suggestion that there is value on ``both sides'' \cite{r1,r4}. However, we believe such views to be in the minority, and that many support work towards a shared vision so long as it is accompanied by certain reasonable concessions. For example, some in Ethics wish to see Alignment acknowledge that talking about existential risk --- regardless of the validity of that concern --- benefits technology companies and thus partially enables the harms they cause \cite{hine2023}.

        \item To the extent that there is a common goal of promoting the responsible development of AI, researchers of all inclinations would benefit from tighter collaborations between members of different groups. The perception of a binary risks debates over AI being co-opted by partisan politics, leading to regulatory deadlocks and acrimonious culture wars on digital platforms. This situation would not benefit any party. Counterproductive antagonism merely creates noise and detracts from actually addressing the many substantive issues.

        \item Given the large overlaps between the two camps, it is likely that progress in Ethics constitutes progress in Alignment, and vice versa. Collaboration across the binary would allow such synergies to emerge.

        \item Lastly, overcoming the perception of a dichotomy may create space for other important discussions which don't fit neatly on either camp's agenda and are currently underrepresented in public debate. For example, the importance of guaranteeing equitable and responsible access to frontier technologies at a global scale, so as to prevent this generation of the technology to increase existing inequality.
        
    \end{itemize}
        
\subsection{How to build bridges}

    \begin{enumerate}
        \item \textbf{Build holistic institutions.}
        \begin{itemize}
            \item Differences over priorities can be resolved organizationally, with different teams or departments (e.g., within a government) assigned responsibility for different concerns, operationalizing holistic normative frameworks.

            \item Examples include: the AI governance roadmap of \textcite{allen2024}, which describes how existing regulatory frameworks and agencies can be empowered and built upon to address the many governance challenges relating to AI.

        \end{itemize}
        
        \item \textbf{Conduct broad church collaborations.}
        \begin{itemize}
            \item Bringing together diverse stakeholders from various disciplines and backgrounds can help foster a more comprehensive understanding of AI's impacts. Such bridge-building can also support the development of inclusive solutions that address the full spectrum of concerns --- ranging from practical matters (e.g., resource allocation) to critiques of the moral purity of particular research agendas --- while acknowledging that distinct kinds of issues call for different responses and methodologies.

            \item Examples include: the policy recommendations produced by the AI Policy and Governance Working Group at the Institute for Advanced Study \cite{ias2023,ias2023a}; open letters calling general-purpose AI to be included in EU AI Act \cite{ainow2023}, restrictions on non-consensual deepfakes \cite{fli2024} and safe harbor for independent AI evaluation \cite{longpre2024}; and more generally, efforts to build a field around Sociotechnical AI Safety \cite{lazar2023a,stanford2023,anwar2024,chen2024}.
        \end{itemize}

        \item \textbf{Where possible, test contentious claims empirically.}
        \begin{itemize}
            \item The more reliable evidence there is, the less space there is for disagreement. Many of the contested claims can be evaluated empirically, and producing empirical answers can create a source of common ground.
            \item Examples include: attempts to assess whether Ethics and Alignment compete for resources and attention \cite{friedrich2023,grunewald2023}; attempts to empirically measure the degree to which current AI systems increase risks such as bio-terrorism \cite{mouton2024}; and adversarial collaborations to identify assumptions or ``cruxes'' that could be tested empirically \cite{rosenberg2024}.
        \end{itemize}
        
        \item \textbf{Be a surprising validator.}
        \begin{itemize}
            \item When people act as surprising validators --- calling out when they actually agree with the ``other side,'' or disagree with their ingroup --- this helps subvert the false binary and avoid pluralistic ignorance (where everyone is afraid to speak up because they feel they are the only person in their community who thinks differently). Such moves can also contribute to the important work of teasing apart good and bad ideas (see \hyperref[sec:complication-3]{above}).
            
            \item Examples include: Ethics acknowledging that not all recent AI hype is snake oil \cite{r5}; or Alignment criticizing those who are dismissive of non-extinction algorithmic harms \cite{critch2023a} or too narrowly technical \cite{critch2024a}, and encouraging people to stop talking about ``p(doom)'' --- a common term among researchers in Alignment that has been criticized by people across the spectrum for being subjective, alarmist, and unscientific \cite{king2024}.
        \end{itemize}
        
        \item \textbf{Where appropriate, develop processes for democratic governance.}
        \begin{itemize}
            \item In some cases, new processes for democratic governance and oversight might help ensure AI systems are responsive to diverse concerns and mitigate risks of power concentrating in any single community.
            \item Examples include: the \textit{Alignment Assemblies} organized by The Collective Intelligence Project \cite{siddarth2023,ganguli2023,cip2023} and similar civil society public-input initiatives \cite{davies2024,hong2024,burgerrat2024}; the projects undertaken as part of the \textit{Democratic Inputs to AI} grant program funded by OpenAI \cite{zaremba2023,openai2023,eloundou2024,perrigo2024}; Meta's announced \textit{Community Forum} on generative AI \cite{clegg2023}; the work of the new \textit{AI \& Democracy Foundation} \cite{aidem2023} to facilitate a community of practice around democratic oversight of AI \cite{ovadya2023,ovadya2023b,ovadya2023d,ovadya2023e}; and efforts to apply insights from social choice theory in AI development \cite{lambert2024a}.
        \end{itemize}
    \end{enumerate}

\section*{Conclusion}
     
    We wrote this piece because we think it is important for people to understand the context and reasons for tension in the community of people working on AI development and governance. But we think it is much \textit{more} important that people do not slip into thinking in the terms of the false binary, seek proactively to build bridges where possible, and try to ensure that the design and governance of AI technology reflects a sufficiently broad range of pluralistic perspectives.

    Optimistically, the current disagreement over what constitutes responsible AI development can be seen as a case of value pluralism, potentially bringing attention to a variety of important concerns. However, there is a risk that this debate is hijacked by culture wars and partisan politics, which could impede meaningful progress on any front. Crucially, researchers across the spectrum concur on the dangers of designing technology without accounting for and mitigating substantial risks, as this could create a difficult-to-reverse path dependency. This shared recognition underscores the urgency of focusing attention and resources on the pressing and substantive concerns raised by all parties involved.

    Despite stemming from different viewpoints and value commitments, all camps in this debate have significant common ground, particularly on fundamental issues. This overlapping consensus provides a strong foundation for productive dialogue. Crucially, as researchers advocate for prioritizing specific issues, they should strive to offer each other terms for cooperation that can be reasonably endorsed by all parties. This approach of seeking mutually acceptable grounds can help bridge ideological divides and facilitate joint efforts to holistically address the complex challenges posed by AI.

\newpage

\section*{Glossary}
\label{sec:glossary}

    Below, we provide loose definitions for some commonly used terms. We emphasize that many of these terms are used differently in different communities and the definitions are contested. Where possible, we have included citations to relevant literature aligned with the Ethics and Alignment perspectives.

    \vspace{3mm}
    {\renewcommand{\arraystretch}{1.5}
    \begin{tabular}{@{}p{33mm}p{93mm}@{}}

        \textbf{``alignment''}
            
            & The degree to which an AI system does what we want it to do, where the definition of ``what we want'' is contested. See \cite{gabriel2020}. \\ 
         
        \textbf{``bias''}
            & Refers broadly to patterns in the behavior of AI systems that disproportionately favor or disfavor specific individuals or groups of people. There are many ways these patterns can be quantified, and some methods for mitigating them. See \cite{danks2017,kordzadeh2022}. \\ 
            
        \textbf{``control problem''}
            & The fact that it's really hard to get an AI system to do what we want it to do. This problem is due in part to the difficulty of specifying ``what we want'' with sufficient richness to account for all possible scenarios and prevent all possible misinterpretations. See \cite{christian2020}.
            \\ 
            
        \textbf{``effective altruism''}
            & A community, strand of moral philosophy, and academic field focused on doing the most good that one can with available resources \cite{singer2015}. \\ 

        \textbf{``existential risk''}
            & Used by Alignment to refer to the risk of human extinction. The word ``existential'' is used more broadly by Ethics to refer to risks that are existential for particular individuals or communities \cite{ord2020}. \\ 
            
        \textbf{``fairness''}
            & Refers broadly to the extent to which harmful bias has been mitigated in the behavior of an AI system, or in how an AI system is used. There are many ways this can be defined and quantified, and some methods for improving it. See \cite{mitchell2021}. \\ %

         \textbf{``longtermism''}
            & Loosely, the view that a human alive in the far future should receive equal weight in moral considerations as a human alive today. See \cite{macaskill2022}. \\
         
        \textbf{``privacy''}
            & Refers broadly to the loss of individual control over personal information when it is used to train AI systems. There are many ways this can be defined and adjudicated in different contexts. See \cite{nissenbaum2004,veliz2021}.

    \end{tabular}}

\newpage
\section*{Acknowledgements}

    Thank you to the many people --- Ethics, Alignment, both, and neither --- who have provided input and feedback on this document. Due to the sensitivity of the topic, we have decided not to acknowledge them by name. Any remaining errors or limitations remain those of the authors.

    Luke Thorburn was supported in part by UK Research and Innovation [grant number EP/S023356/1], in the UKRI Centre for Doctoral Training in Safe and Trusted Artificial Intelligence (\href{https://safeandtrustedai.org}{safeandtrustedai.org}), King's College London.

\section*{References}

    Due to the sensitivity of the topic, we have decided to withhold some references in order to (a) protect individuals and (b) focus attention on the overall conflict dynamics, rather than one-off incidents. If you need to access any of the withheld references, please contact the authors.

    \printbibliography[heading=none]

\end{document}